# Optimization of a Locally Shielded 16-Channel Transmit Array for Human Brain MRI at 10.5 T via Electromagnetic/Circuit Co-Simulation and Parallel Transmission Strategies

Mengxuan Zheng[1], Xin Shao[2], Matt Waks[2], Mazin M. Mustafa[2], Kamil Ugurbil[2], Xiaoping Wu[2], Xiaotong Zhang[3], Alireza Sadeghi-Tarakameh[2]

***Abstract*—Modern neuroimaging and neuroscience increasingly demand high spatial and temporal resolution, driving the use of ultrahigh-field ( ≥ 7 T) MRI for improved signal-to-noise ratio. Realizing this advantage depends critically on advanced radiofrequency (RF) and gradient technologies, particularly multi-channel transmit arrays and high-performance head-gradient systems. In integrated systems, a local RF shield must provide sufficient field confinement while preserving transmit performance and minimizing sensitivity to nearby conductive structures. In this work, we sought to optimize a locally shielded 16-channel transmit RF array for human neuroimaging at 10.5 T using electromagnetic/circuit co-simulation combined with parallel-transmission strategies. Candidate transmit-array and RF-shield configurations were evaluated before and after gradient-shield integration to identify a configuration with suitable RF transmit and shielding performance for the integrated system. The results show that no single standalone metric, including static RF shimming performance, dynamic parallel transmission (pTx) performance, or external-field leakage, reliably predicts robustness after gradient-shield integration. Progressive screening identified the aligned loop-folded dipole (LFD) array combined with a compact cylindrical shield featuring a conical top and three windows (160-mm radius and 70-mm window length) as the final configuration. Without RF-circuit retuning, the selected design maintained external-field leakage below 1% and after gradient-shield integration achieved whole-brain excitation coefficient of variation (CoV) of approximately 27% with static RF shimming and 9% with dynamic pTx. Experimental validation of the unshielded LFD array showed good agreement between simulated and measured B1+ efficiency distributions, with a normalized root-mean-square error (NRMSE) of approximately 8%.**

This work was supported in part by the National Natural Science Foundation of China under Grant 52277232 and Grant 52293424; and in part by USA NIH under Grants R01 NS136490, R01 EB038654, and P41 EB027061.
*(Corresponding author: Xiaoping Wu and Xiaotong Zhang).*
Mengxuan Zheng is with the College of Information Science and Engineering, Northeastern University, Shenyang 110819, China (e-mail: zhengmengxuan@ise.neu.edu.cn).
Xiaoping Wu, Xin Shao, Matt Waks, Mazin Mustafa, Kamil Ugurbil, and Alireza Sadeghi-Tarakameh are with the Center for Magnetic Resonance Research, Department of Radiology, University of Minnesota, Minneapolis, MN 55455 USA (e-mail: wuxxx184@umn.edu)
Xiaotong Zhang is with the College of Electrical Engineering, Zhejiang University, Hangzhou 310027, China (e-mail: zhangxiaotong@zju.edu.cn).
Supplementary material for this article is provided by the authors.

## I. INTRODUCTION

ULTRA high-field (UHF; ≥ 7T) MRI offers inherently higher signal-to-noise ratio (SNR)[1], [2], [3], [4], which can be leveraged to achieve higher spatial resolution or shorter scan times or both. These advantages have enabled a broad range of anatomical[5], [6], [7], [8] and functional [9], [10], [11], [12] neuroimaging applications. The substantially higher SNR can exhibit superlinear scaling with increasing field strength[4], [13], [14], [15], [16], [17]. However, these benefits are accompanied by pronounced radiofrequency (RF) related challenges, including severe transmit magnetic field (B1+) nonuniformity[18], elevated local specific absorption rate (SAR)[19], [20], [21], and limited peak transmit efficiency[22], [23]. The recently realized beyond-10T regime[24], [25], particularly early human brain studies at 10.5 T[11], [16], [17], [26], [27], [28], [29], [30], [31], has already demonstrated both the promises and challenges at this UHF. A principal approach to addressing these challenges is the use of multi-channel transmit RF arrays combined with parallel transmission (pTx)[32], [33], [34], which requires appropriate array design, RF optimization, and safety validation[24], [27], [35], [36]. At the same time, the electromagnetic (EM) environment surrounding UHF RF coils has increasingly been treated as a design variable, with RF shields and other passive structures shown to alter transmit-field distribution, coupling, and B1+ efficiency[37], [38], [39], [40].

Transmit RF array design and management therefore become even more demanding in modern neuroimaging platforms that incorporate high-performance head gradient coils[41], where the RF array operates in close proximity to additional conductive structures. In such systems, local RF shielding may be required to mitigate electromagnetic interference, reduce undesired radiation, and satisfy system-integration constraints. Electromagnetically, however, the RF shield also forms part of the boundary condition experienced by the transmit array, and its geometry can therefore alter the array response. Chen *et al.* showed that the surface-impedance characteristics of a shielding plate can substantially alter the electric- and magnetic-field distributions of a 7-T MRI dipole element, demonstrating that the shielding structure directly participates in determining RF field behavior[42]. More generally, shield geometry can influence inter-element coupling[43], impedance behavior, radiation loss, and field distributions within the sample, thereby affecting achievable transmit performance[44], [45]. These observations motivate array-level evaluation of shielded transmit arrays within their intended electromagnetic environment, with the shield considered an integral part of the RF system.

Recent studies have increasingly treated the RF shield as an integral electromagnetic component of UHF head-coil design. Shield geometry has been shown to influence array decoupling, field confinement, B1+ efficiency, and receive SNR across UHF systems[43], [45], [46]. At 10.5 T, prior numerical work has mainly focused on RF-shield effects on receive-array SNR, whereas the influence of the surrounding gradient structure on shielded transmit-array behavior remains less explored[46]. Transmit-array evaluations at UHF have commonly relied on static RF shimming or selected excitation scenarios[33], [47], [48], [49], including recent 10.5-T head-array studies[16], [17], [26], [27], [29] and universal-shim demonstrations[50], [51]. Beyond such static evaluations, SAR-constrained pTx pulse design provides a more application-relevant means of evaluating array performance under realistic excitation and safety constraints[52], [53], [54], [55], [56], while simulation-based array comparisons have further incorporated excitation homogeneity, SAR, and B1+ efficiency metrics[57], [58]. For shielded transmit arrays, Chu *et al.* [45], [59] further demonstrated that EM simulation combined with pTx pulse design can reveal excitation-fidelity and local SAR tradeoffs that are not captured by isolated B1+ metrics alone.

Despite these advances, shielded UHF transmit-array design remains a multifactor problem, particularly for compact neuroimaging platforms incorporating high-performance local gradient hardware. In such systems, the RF shield must simultaneously satisfy requirements for transmit performance, field confinement, accessibility, and hardware integration, while the nearby gradient structure introduces an additional conductive boundary that can perturb array loading and multiport impedance. The extent of this perturbation may depend strongly on RF-shield geometry, such that configurations with similar standalone transmit and shielding performance can exhibit substantially different robustness after gradient integration. Whether conventional transmit-performance and leakage metrics can predict this sensitivity remains unclear. This motivates evaluating candidate shield geometries not only for transmit and shielding performance, but also for susceptibility to gradient-induced loading and impedance perturbations.

In this study, we developed a computational framework to design, optimize and evaluate a locally shielded 16-channel transmit array for human brain MRI at 10.5 T. The framework integrates full-wave EM simulation, EM/circuit co-simulation, SAR-aware static RF shimming and dynamic pTx pulse design, RF-field leakage assessment, and evaluation under the intended local-gradient environment. Multiple array-shield configurations were progressively screened to identify the design that balanced transmit performance and shielding effectiveness within the visual-access and gradient-integration constraints of the target platform. The final selection favored configurations that retained RF performance and shielding effectiveness while reducing sensitivity to the surrounding conductive environment. The resulting analysis provides practical design guidance for shielded transmit RF arrays intended for integration with high-performance head-gradient systems at 10.5 T.

## II. Methods

### A. Workflow Overview

The proposed computational framework (Fig. 1) integrates four main components: EM/circuit co-simulation, local SAR and virtual observation points (VOPs) calculation, SAR-aware transmit performance evaluation, and RF and gradient shield integration assessment. It was used to compare transmit-array–RF-shield configurations for 10.5 T human brain imaging.

For each array–shield configuration, port-wise S-parameters and EM fields were first obtained from full-wave simulations. Circuit-level tuning, matching, and decoupling were subsequently performed through EM/circuit co-simulation [60]. The resulting network responses were used to synthesize multi-channel magnetic and electric fields for arbitrary complex excitation weights.

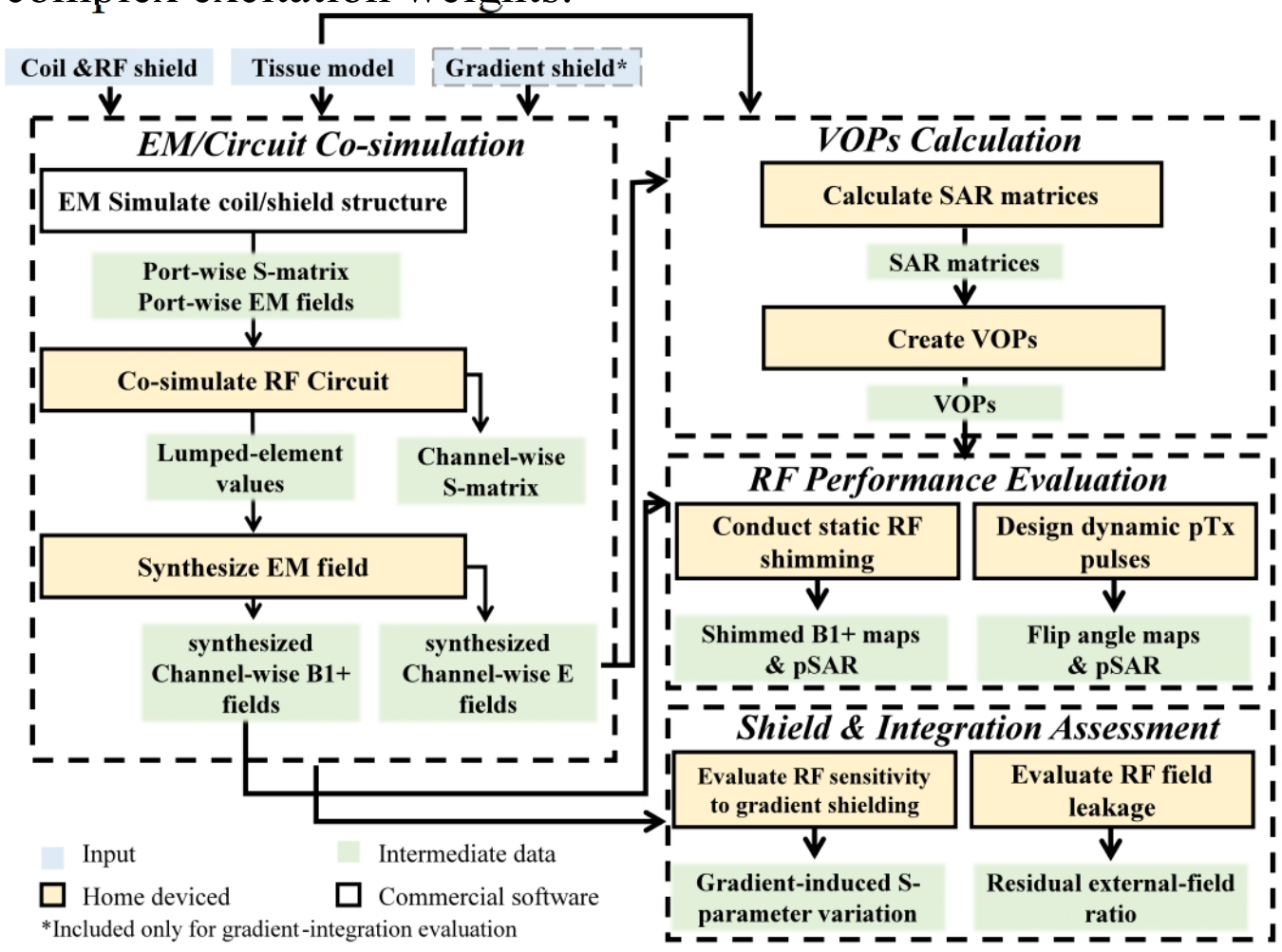

**Fig. 1.** Proposed computational framework for the design and evaluation of RF transmit arrays. The framework consists of four modules: 1) EM/circuit co-simulation for array tuning, matching, decoupling, and multi-channel EM-field synthesis, 2) Virtual observation points (VOPs) calculation, including local SAR-matrix calculation and VOP compression, 3) RF transmit-performance evaluation using SAR-aware static RF shimming and dynamic pTx pulse design, and 4) shield and integration assessment based on RF-field leakage and sensitivity to gradient-shield integration. Coil geometry, RF-shield geometry, and the tissue model serve as common inputs. The gradient shield is introduced only for the integration analysis.

Voxel-wise local SAR matrices were then constructed from the electric fields and compressed into VOPs[61] for efficient incorporation of local SAR constraints into pTx optimization. Transmit performance was evaluated using SAR-constrained static RF shimming as the general geometry-comparison approach and SAR-constrained spectral-spatial (SPSP) dynamic pTx pulse design[62] as a complementary application-specific evaluation.

In addition to transmit performance, the framework evaluates hardware-integration effects that are not captured by conventional field-efficiency metrics alone. RF-field leakage through the shield is quantified, and the sensitivity of RF performance to gradient-shield integration is assessed by introducing the gradient shield only in the corresponding integration analysis. Coil geometry, RF-shield geometry, and the tissue model are otherwise maintained as common inputs across the computational workflow.

All postprocessing, optimization, SAR analysis, and pTx pulse calculations were performed in MATLAB. Full-wave 3-D EM simulations were performed in ANSYS HFSS (Ansys Inc., Canonsburg, PA, USA).

### *B. Transmit-Array and RF-Shield Configurations*

Three 16-channel transmit-array topologies were evaluated: self-decoupled loop (SD) arrays with staggered and aligned row layouts, and an aligned loop-folded dipole (LFD) array (Fig. 2). Each SD element was modeled with six ports: one physical transmit port, two matching ports, and three tuning ports, yielding 96 ports for the 16-channel array. Each LFD element was modeled with three ports: one physical transmit port and two lumped ports for tuning and matching, yielding 48 ports in total.

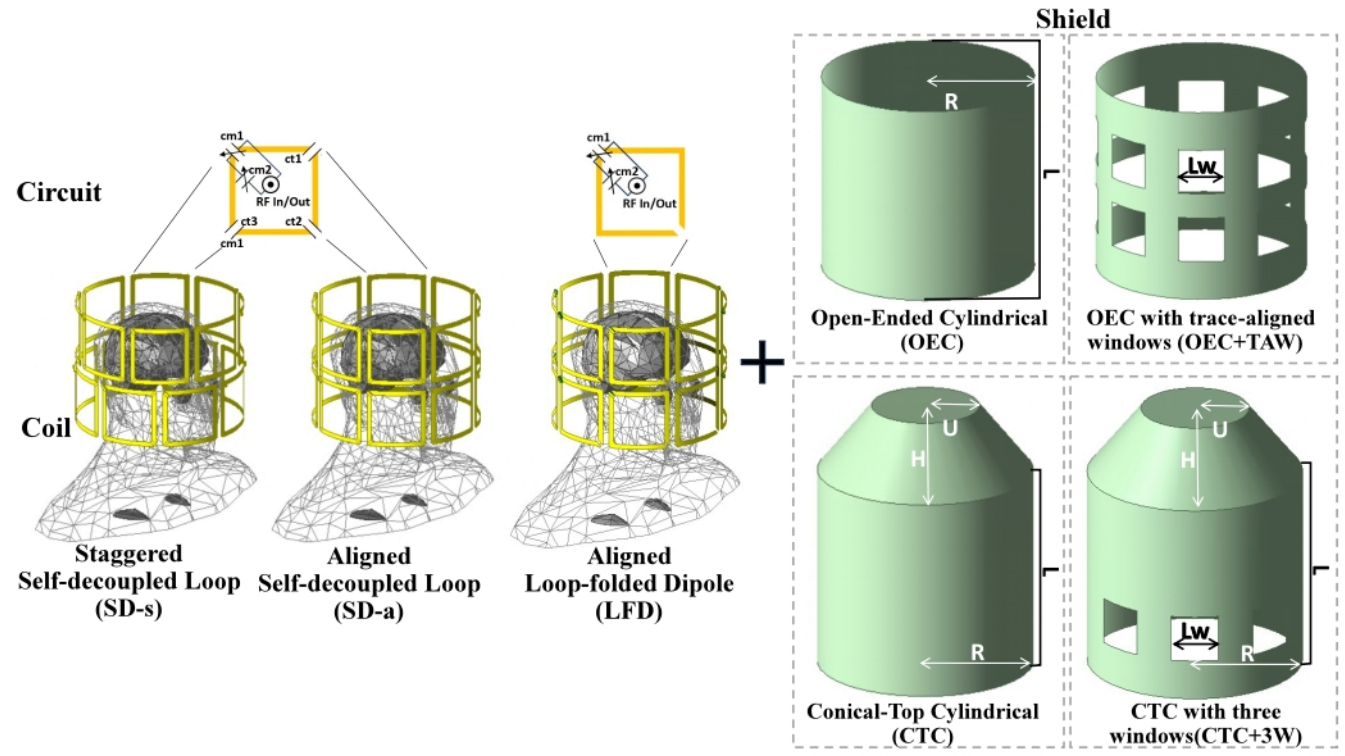

**Fig. 2**. Geometries of the 16-channel dual-row transmit arrays and RF shields studied. The unshielded transmit arrays (left panel) include staggered-row and aligned-row self-decoupled loop arrays and an aligned loop-folded dipole array, with associated single-element circuit models shown on top. The RF shields (right panel) include the open-ended cylindrical (OEC), OEC with trace-aligned windows (OEC+TAW), conical-top cylindrical (CTC), and CTC with three windows (CTC+3W) configurations.

Four RF-shield geometries (Fig. 2) were investigated: open-ended cylindrical (OEC), OEC with trace-aligned windows (OEC+TAW), conical-top cylindrical (CTC), and CTC with three windows (CTC+3W). To accommodate visual access and experimental hardware integration in task-based fMRI applications, the OEC+TAW and CTC+3W configurations incorporated open-access features. The OEC+TAW geometry used distributed openings, whereas the CTC+3W geometry introduced three localized windows positioned anterior to the second array row and directly in front of the subject's face. These two configurations were included to assess the tradeoff among transmit performance, RF-field confinement, and physical accessibility.

The geometric parameters and evaluated values are summarized in Supplementary Fig. S1 and Table S1. A common dual-row array arrangement, loaded head model, and brain evaluation mask were maintained across all configurations to enable consistent comparison.

### *C. EM/Circuit Co-Simulation*

*1) Full-Wave 3-D EM Simulation:* Full-wave EM simulations were performed in ANSYS HFSS using the finite-element method at 447 MHz, corresponding to the proton Larmor frequency at 10.5 T. The coil conductors and RF shields were modeled as perfect electric conductors. A male head-and-shoulder model with 4-mm isotropic resolution was used as the loading model for all configurations.

Each EM port, including the transmit and lumped-element ports, was excited sequentially while all remaining ports were terminated with 50-Ω loads. Perfectly matched layers were applied at the outer simulation boundary. The complete multiport scattering matrix (S-matrix) and the port-wise EM field bases were extracted at 447 MHz. The field data were exported on a 4-mm isotropic grid. Identical loading position, boundary conditions, and solver settings were maintained across all configurations, with an adaptive convergence tolerance of 0.002.

*2) Network Reduction and Circuit Optimization:* Tuning, matching, and decoupling were implemented by terminating the lumped-element ports of the HFSS-derived multiport network with variable capacitor and inductor impedances. After separating the physical transmit ports from the lumped-element ports, the reduced S-matrix at the physical transmit ports was obtained analytically as[63]:

$$\boldsymbol{S}_{phys} = \boldsymbol{S}_{pp} + \boldsymbol{S}_{pl}\boldsymbol{S}_{lump}\left(I - \boldsymbol{S}_{ll}\boldsymbol{S}_{lump}\right)^{-1}\boldsymbol{S}_{lp}. \quad (1)$$

Here, $\boldsymbol{S}_{pp}$ and $\boldsymbol{S}_{ll}$ denote the scattering submatrices for the physical transmit and lumped-element ports, respectively; $\boldsymbol{S}_{pl}$ and $\boldsymbol{S}_{lp}$ denote scattering submatrices describing mutual coupling between physical transmit and lumped-element ports, respectively; $\boldsymbol{S}_{lump}$ is the diagonal reflection matrix associated with the lumped-element terminations. Capacitors and inductors were modeled with a 0.1-Ω series resistance to account for component loss.

The lumped-element values were optimized using a two-stage global-to-local procedure. In the first stage, CMA-ES[64] was used for global exploration by solving:

$$\min_{\boldsymbol{x}} J_{\text{CMA}}(\boldsymbol{x}),$$

$$\text{Where: } J_{\text{CMA}}(\boldsymbol{x}) = \frac{\left\| max(0,\ \left|\boldsymbol{S}_{phys,ii}(\boldsymbol{x}) - \gamma_m\right| \right\|_2}{\gamma_m} + \lambda \frac{\left\| max(0,\ \max_{i\neq j}\left|\boldsymbol{S}_{phys,\,ij}(\boldsymbol{x})\right| - \gamma_c) \right\|_2}{\gamma_c}. \quad (2)$$

Where $\boldsymbol{x}$ denotes the lumped-element values, and $\gamma_m = 10^{-15/20}$, $\gamma_c = 10^{-10/20}$, correspond to the −15 dB matching and −10 dB inter-channel coupling criteria, respectively, and $\lambda$ controls their relative weighting. Candidates satisfying both criteria were prioritized according to their worst inter-channel coupling. The best feasible CMA-ES solution was subsequently refined using MATLAB *fmincon*:

$$\min_{\boldsymbol{x}} \max_{i \neq j} \left| S_{\mathrm{phys},ij}(\boldsymbol{x}) \right|,$$
$$\text{subject to: } \left| \boldsymbol{S}_{\mathrm{phys},ii}(\mathbf{x}) \right| \le \gamma_m,\ \forall i,$$
$$\left| \boldsymbol{S}_{\mathrm{phys},ij}(\mathbf{x}) \right| \le \gamma_c,\ \forall i \neq j. \quad (3)$$

Thus, the global stage provided broad exploration of the matching-coupling tradeoff, whereas the local stage minimized the worst inter-channel coupling within the feasible matching region.

*3) Multi-Channel EM Field Synthesis:* For a given set of optimized lumped-element values, the physical-channel S-matrix was derived and the corresponding channel-wise EM fields were synthesized from the multiport EM simulation using the circuit co-simulation formulation described by Beqiri et al. [63]. The B1+ field was then calculated from the transverse magnetic-field components as[65]:

$$B_1^+ = \left(B_x + jB_y\right)/2. \quad (4)$$

To synthesize the physical channels, each of the 16 transmit channels was mathematically excited with a 1-V peak incident wave, while the other 15 transmit ports were set to zero. The resulting channel-specific EM field maps were referenced to a 50-Ω port impedance and were used to inform subsequent RF-shimming, pTx pulse design, and SAR calculations.

### D. Local SAR Matrices Calculation and VOP Compression

For each array–shield configuration, local SAR matrices were calculated from the 16-channel electric-field maps of the physical transmit channels. Voxel-wise SAR matrices were first constructed from the channel-wise electric fields together with the corresponding tissue conductivity and mass density, and were subsequently mass-averaged over 10 g of contiguous tissue to obtain the 10-g local SAR matrices[66].

The 10-g SAR matrices were compressed into a reduced set of VOPs using a 10% overestimation tolerance[48]. The resulting VOPs were subsequently used to impose peak 10-g local SAR constraints during SAR-aware RF-shimming and pTx optimization. The same tissue model, electrical properties, 10-g averaging procedure, and VOP compression tolerance were applied to all array–shield configurations.

### E. RF Performance Evaluation

RF performance was evaluated using SAR-aware pTx strategies including both static RF shimming and dynamic pTx pulse design. Static RF shimming was used for geometry comparison across excitation uniformity, B1+ efficiency, and SAR efficiency, whereas dynamic pTx was used for application-specific evaluation under a common pulse-design protocol.

*1) Static RF Shimming:* For each array-shield configuration, static RF shimming was performed using MATLAB *fmincon*. The complex shim vector was optimized to minimize whole-brain B1+ NRMSE subject to a prescribed B1+ efficiency target and a VOP-based peak 10-g local SAR constraint. Sweeping the efficiency and SAR constraints generated a three-dimensional tradeoff among excitation uniformity, B1+ efficiency, and local SAR, as illustrated in Supplementary Fig. S2. For the cross-configuration comparison in the main text, the peak local SAR constraint was fixed at 0.4 W/kg, referenced to the 1-V peak incident-wave basis used for channel-field synthesis. This value served as a common optimization budget rather than a physiological SAR limit. The B1+ efficiency target was swept from 0.8 to 1.25 times the corresponding CP-mode efficiency, and the minimum-NRMSE feasible solution was retained at each target.

Transmit performance was subsequently quantified using CoV, NRMSE, B1+ efficiency, and SAR efficiency:

$$\mathrm{CoV} = \frac{\sqrt{\frac{1}{N}\sum_{r\in\mathrm{ROI}}\left(|B_1^+(r)| - \frac{1}{N}\sum_{r\in\mathrm{ROI}}|B_1^+(r)|\right)^2}}{\frac{1}{N}\sum_{r\in\mathrm{ROI}}|B_1^+(r)|}, \quad (5)$$

$$\mathrm{NRMSE} = \frac{\sqrt{\frac{1}{N}\sum_{r\in\mathrm{ROI}}\left(|B_1^+(r)| - B_{1,\mathrm{target}}^+(r)\right)^2}}{\frac{1}{N}\sum_{r\in\mathrm{ROI}} B_{1,\mathrm{target}}^+(r)} \times 100\%, \quad (6)$$

$$\eta_{B_1^+} = \frac{\frac{1}{N}\sum_{r\in\mathrm{ROI}}|B_1^+(r)|}{\sqrt{P_{\mathrm{in}}}},\ \eta_{\mathrm{SAR}} = \frac{\frac{1}{N}\sum_{r\in\mathrm{ROI}}|B_1^+(r)|}{\sqrt{\mathrm{pSAR_{10g}}}}. \quad (7)$$

Here, $N$ denotes the number of voxels within the whole-brain ROI, $P_{\mathrm{in}}$ is the total incident RF power and $\mathrm{pSAR_{10g}}$ is the peak 10-g local SAR estimated using VOPs.

A representative shim was selected from the feasible solutions of each configuration using a two-objective corner-distance criterion favoring lower NRMSE and higher SAR efficiency. NRMSE and SAR efficiency were first min-max normalized within each candidate set, and the weighted distance to the ideal low-NRMSE/high-SAR-efficiency corner was calculated as

$$D = \sqrt{(\lambda_1 \mathrm{NRMSE_{norm}})^2 + \left[\lambda_2\left(1 - \eta_{\mathrm{SAR,norm}}\right)\right]^2}. \quad (8)$$

Here, $\lambda_1 = 2$ and $\lambda_2 = 1$ were used.

Among the highest-ranked candidates, solutions exhibiting pronounced localized low-B1+ regions (“black holes”) were excluded before one representative excitation was retained for each configuration.

*2) Dynamic pTx Pulse Design:* Dynamic pTx pulse design was performed to further evaluate transmit performance. In particular, SAR-constrained 4D spectral-spatial pTx pulses [56] were designed to achieve uniform water-selective excitation across the brain. The pulse-design inputs consisted of the multi-channel B1+ maps, compressed VOP matrices, and the whole-brain ROI. In all cases, identical pulse-design settings were used assuming no B0 inhomogeneities.

Pulses were designed with a total duration of ~2.8 ms using 24 kT-points to achieve 13° nominal flip angles. The spectral separation between water and fat was set according to the chemical shift at 10.5 T. Pulses were optimized subject to a peak 10-g local SAR constraint of 10 W/kg. CoV of flip angle distribution across the entire brain was then calculated to evaluate pulse performance.

### F. RF Shield and Integration Evaluation

*1) RF Field Leakage Evaluation*: In this study, RF-shielding effectiveness was used as a measure of RF-field confinement and was quantified using the external RF-field leakage metric defined below. To characterize the RF field extending beyond the RF shield and its potential interaction with the nearby gradient shield, external RF-field leakage was quantified under CP-mode excitation using a residual electric-field ratio, defined as the mean voxel-wise electric-field magnitude within an external evaluation region normalized by that within an internal reference region. The internal reference region comprised the air space between the head and the RF shield, spanning 160 mm in the head-foot direction. The primary leakage ratio was evaluated in the nearest external region (R1), an annular zone extending 32 mm radially outward from the RF shield. Residual-field distributions at larger distances are reported in Supplementary Fig. S3.

*2) RF Sensitivity to Gradient-Shield Integration*: RF sensitivity to gradient-shield integration was evaluated by including a cylindrical gradient shield (of 300 mm in length) in a separate HFSS model for each array–shield configuration. Gradient-shield radii of 210, 220, and 230 mm were evaluated. For each integrated geometry, the resulting full multiport S matrix was reduced to the 16 physical transmit ports using the same lumped-element network optimized for the corresponding configuration without the gradient shield.

Gradient-shield-induced RF perturbation was evaluated by comparing the baseline configuration without the gradient shield (G0) with the corresponding configuration after gradient-shield integration while retaining the G0-optimized lumped-element network (G1-retained). This fixed-network comparison was used to assess the intrinsic sensitivity of each RF-shield geometry to the added conductive boundary. Relevant metrics included the worst-channel reflection coefficient, worst inter-channel coupling, and relative S-matrix variation:

$$D_S = \frac{|S_{\mathrm{G1}} - S_{\mathrm{G0}}|_F}{|S_{\mathrm{G0}}|_F} \tag{9}$$

where $S_{\mathrm{G0}}$ and $S_{\mathrm{G1}}$ denote the channel-wise S matrices without and with the gradient shield, respectively, and $|\cdot|_F$ denotes the Frobenius norm. These metrics characterize changes in impedance matching, inter-channel coupling, and overall multiport network response to the gradient shield.

To further characterize gradient-shield-induced EM loading, the full 48-port HFSS impedance matrices were compared. The gradient-shield-induced change in self-reactance at each port was defined as:

$$|\Delta X_{ii}| = \left|\mathrm{Im}\left(Z_{ii}^{G1} - Z_{ii}^{G0}\right)\right| \tag{10}$$

where $Z_{ii}^{G0}$ and $Z_{ii}^{G1}$ denote the self-impedance of port $i$ without and with the gradient shield, respectively, and $\mathrm{Im}(\cdot)$ means the imaginary part of a complex number. The ports were grouped according to their roles in the circuit model as drive-associated, capacitor-associated, and inductor-associated ports.

*3) Post-Gradient-Shield-Integration RF Performance Evaluation:* For the selected RF-shield designs, post-integration transmit performance was evaluated without RF-circuit retuning, using the G0-optimized lumped-element values. For the G1-retained geometry, the port-wise EM field bases were regenerated and combined with the fixed circuit network to synthesize the corresponding 16-channel EM field bases. The 10-g local SAR matrices and VOP representation were correspondingly recalculated for the integrated geometry. Static RF shimming and dynamic pTx pulses were independently reoptimized using the G1 channel-wise field bases, with the same settings as in Sections II-E1 and II-E2. The resulting excitation metrics were compared with their G0 counterparts to quantify the robustness of transmit performance to gradient-shield integration without RF-circuit retuning.

### G. Experimental Validation

The B1+ field prediction obtained from the EM/circuit co-simulation framework was experimentally evaluated using the unshielded 16-channel LFD array under CP-mode excitation. A lightbulb-shaped uniform phantom was filled with a tissue-mimicking polyvinylpyrrolidone (PVP) solution with electrical properties of conductivity 0.65 S/m and relative permittivity 47.2. The solution contained 651.1 g/L PVP, 17.83 g/L NaCl, 0.48 g/L $NiCl_2$ $6H_2O$, and 20 g/L agar in deionized water.

Absolute B1+ maps were acquired using the actual flip-angle imaging (AFI) method[67] with $TE/TR_1/TR_2$ = 1.72/28/128 ms, nominal flip angle of 60°, and spatial resolution of 1.5×1.5×5 $mm^3$. For comparison, simulated and measured B1+ maps were normalized by the square root of the corresponding input power to create B1+ efficiency maps. Agreement between simulation and experiment was quantified by calculating NRMSE over the three-dimensional phantom mask.

## III. Results

### A. Unshielded Transmit-Array Topology Screening

The three unshielded 16-channel transmit-array topologies were compared using static RF shimming under identical optimization settings and a common VOPs-based peak 10-g local SAR constraint. The three arrays exhibited comparable whole-brain excitation uniformity but pronounced differences in B1+ and SAR efficiencies (Fig. 3). Relative to the staggered self-decoupled loop array (SD-s), the aligned self-decoupled loop array (SD-a) produced only a modest change in uniformity while increasing both efficiency metrics. The aligned loop-folded dipole (LFD) array further increased efficiencies, yielding ~90% higher B1+ efficiency and ~90% higher SAR efficiency than SD-a, at a slight increase in CoV (~28% vs. ~27% for SD-a).

Dynamic pTx evaluation provided an additional application-specific comparison, with the LFD topology yielding the lowest flip-angle CoV. Corresponding three-view B1+ distributions, flip-angle maps, and flip-angle CoV values are shown in Supplementary Fig. S4.

Thus, the LFD topology was selected for the subsequent RF-shield study.

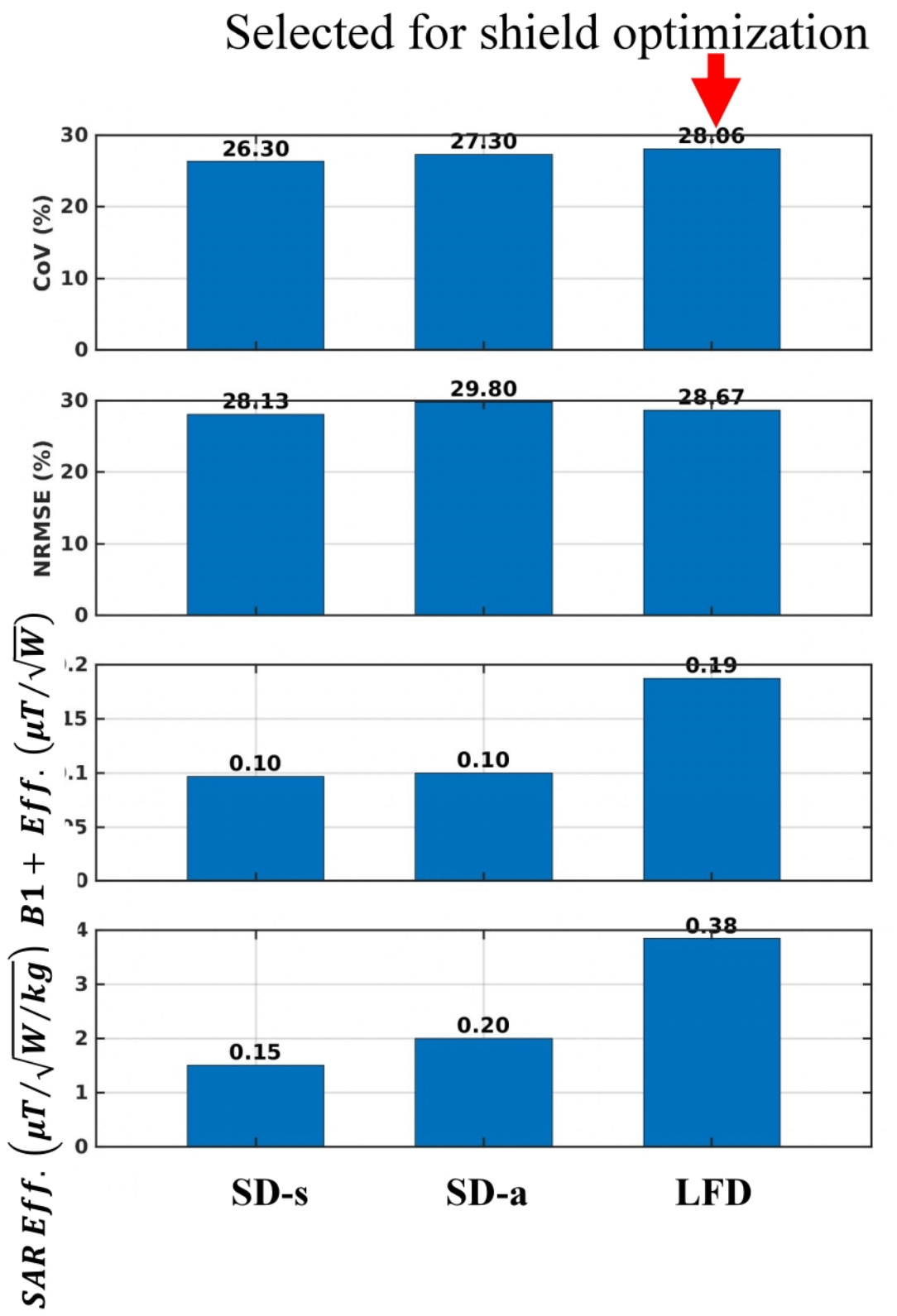

**Fig. 3.** Static RF shimming comparison of the three unshielded 16-channel transmit arrays. Whole brain CoV, B1+ efficiency, and SAR efficiency are reported for SD-s, SD-a, and LFD.

### *B. Geometry-Dependent RF and Shielding Performance*

The RF-shield geometries were subsequently evaluated with the LFD array, with geometry-specific screening based on static RF-shimming performance and CP-mode external-field leakage, followed by dynamic-pTx evaluation of the retained configurations. Fig. 4 summarizes the sequential screening of the OEC and OEC+TAW geometries. For the OEC shield, increasing the shield radius reduced excitation nonuniformity over the tested range, with the 190-mm-radius configuration providing the most uniform excitation among the OEC cases. This improvement was accompanied by increased R1 leakage, indicating a tradeoff between transmit performance and field confinement as the array-to-shield spacing increased. Based on this tradeoff, the 190-mm-radius OEC configuration was retained for cross-geometry comparison.

For the OEC+TAW geometry, shield length, window size, and radius were varied sequentially to reduce the design space. Shortening the shield improved several transmit-performance metrics but caused a pronounced increase in external-field leakage. A shield length of 300 mm was therefore retained to preserve axial field confinement before the window-size and radius sweeps. The subsequent parameter sweeps identified the 70-mm-window, 160-mm-radius configuration as the retained OEC+TAW design. The dependence of performance on window size also varied with the surrounding shield dimensions, showing that the effect of the openings could not be considered independently of the overall shield geometry.

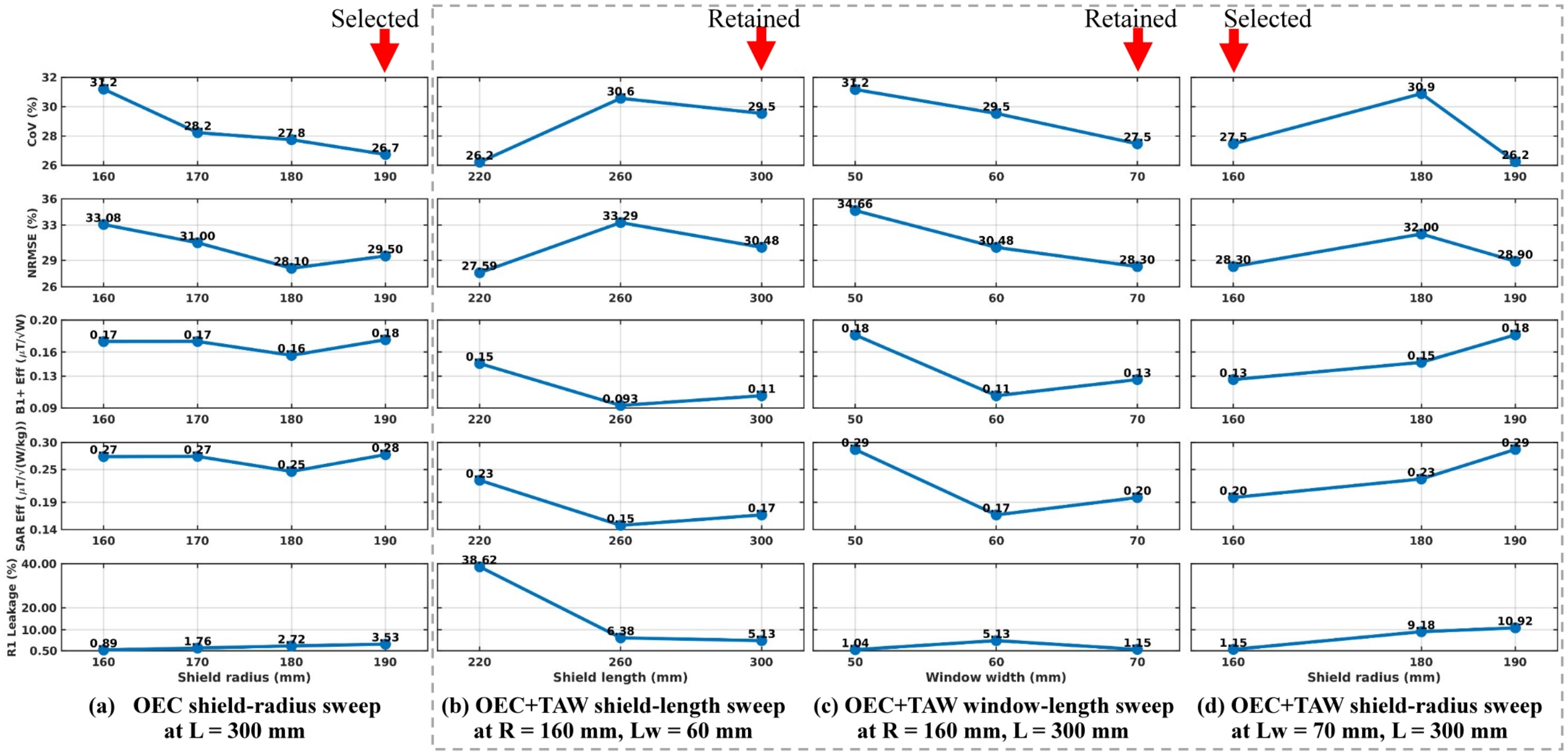

**Fig. 4.** Screening of the OEC and OEC+TAW shield geometries based on static RF shimming performance and CP-mode R1 leakage. (a) OEC shield-radius sweep at fixed shield length of 300 mm (i.e., L=300 mm). OEC+TAW sweeps of (b) shield length (L), (c) window size (Lw), and (d) shield radius (R) were performed sequentially, with the parameter value retained from each prior sweep being used in subsequent sweeps of other parameters. In all cases, R1 leakage ratios (bottom) were calculated using the R1 region defined in the Methods.

The CTC and CTC+3W geometries exhibited a similar geometry-dependent tradeoff between transmit performance and external-field leakage (Fig. 5). For the CTC shield, the cap dimensions were first varied at a fixed shield radius of 160 mm. At a given shield radius and length, the RF performance showed a systematic dependence on cap size. In particular, the U=80 mm,

H=50 mm and U=40 mm, H= 100 mm configurations, which have similar cap volumes, produced similar RF performance. Among the evaluated cases, the H=100 mm, U=80 mm configuration provided a favorable balance between transmit performance and leakage and was retained as the baseline cap geometry for the CTC+3W study.

For CTC+3W, the window dimensions were varied at shield radii of 160 and 190 mm. The window-size dependence differed between the two radii, further demonstrating that the effect of the local openings depended on the overall shield dimensions. Two configurations representing different performance tradeoffs were retained for cross-geometry comparison: R190/W50, which exhibited higher SAR efficiency, and the more compact R160/W70 configuration, which exhibited lower leakage while maintaining comparable excitation uniformity.

Overall, the shield sweeps demonstrated that array-to-shield spacing, axial enclosure, cap geometry, and localized openings jointly determined the balance between transmit performance and RF-field confinement. No individual geometric parameter produced a monotonic improvement in both transmit performance and leakage over the investigated design space.

### *C. Cross-Family RF-Shielding Tradeoff*

Following geometry-specific screening, the retained configurations were compared under common static RF shimming, dynamic-pTx, and leakage-evaluation protocols (Fig. 6). The comparison revealed a clear tradeoff between transmit performance and external-field confinement. The OEC R190 configuration yielded the lowest static-shim CoV (26.7%) and a high SAR efficiency of 0.28, but also exhibited the largest R1 leakage (3.63%). In contrast, the OEC+TAW R160/W70 and CTC H100/U80 configurations reduced leakage to approximately 1%, accompanied by lower SAR efficiency.

The CTC+3W configurations provided a more balanced tradeoff between excitation performance and field confinement. The R190/W50 design maintained relatively high SAR efficiency (0.25) with reduced leakage (1.90%), whereas the more compact R160/W70 design further reduced leakage to 0.84% while preserving comparable excitation uniformity (CoV, 27.1%). These two configurations therefore represented distinct operating points, with R190/W50 favoring RF efficiency and R160/W70 favoring field confinement and compactness.

Dynamic pTx substantially reduced excitation nonuniformity across all retained configurations, with whole-brain CoVs ranging from 6.9% to 8.8% [Fig. 6(b)]. Despite the reduced spread in excitation uniformity, the relative cross-geometry trends remained consistent, and both CTC+3W configurations maintained low nonuniformity under the common pTx protocol.

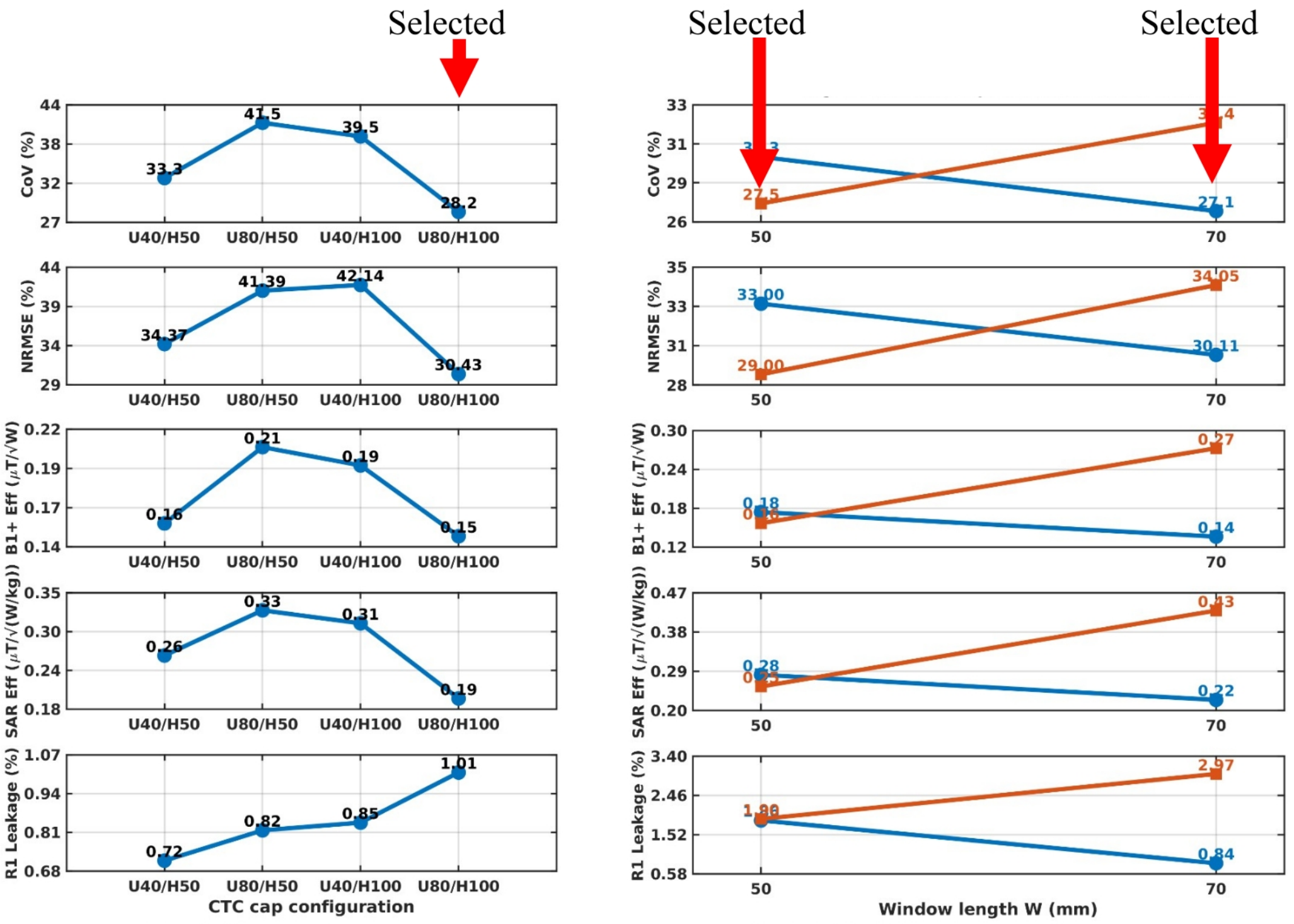

**(a) CTC cap-size sweep at R = 160 mm, L = 300 mm**

**(b) CTC+3W window-size sweep at R = 160/190 mm, H = 100 mm, U/R = 0.5**

**Fig. 5.** Screening of the CTC and CTC+3W shield geometries based on static RF-shimming performance and CP-mode R1 leakage. (a) CTC cap-geometry sweeps of cone-top upper radius (U) and height (H) at fixed cylindrical radius of 160 mm (i.e., R=160 mm). (b) CTC+3W window-size sweep at R=160 and 190 mm using the selected cap geometry from (a) (i.e., H=100 mm, U/R=0.5). In all cases, the cylindrical length was fixed at 300 mm (i.e., L=300 mm).

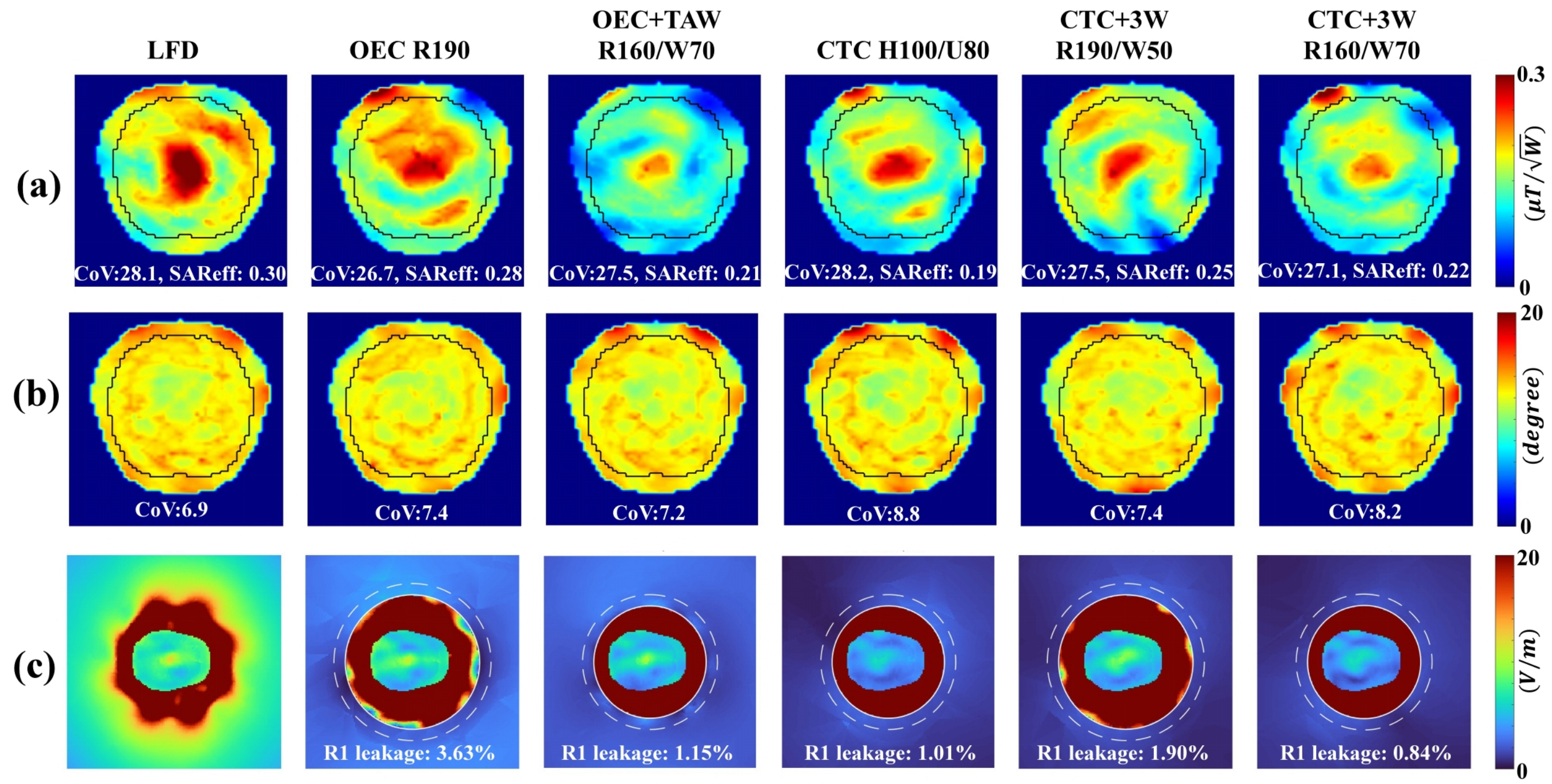

**Fig. 6.** Cross-family comparison of the selected RF-shield configurations together with the unshielded LFD baseline, without gradient-shield integration. (a) Normalized B1+ maps after static RF shimming, with whole-brain CoV and SAR efficiency reported. (b) Flip-angle distributions obtained using SAR-aware dynamic pTx pulse design with whole-brain CoV reported. (c) Large-FOV CP-mode electric-field distributions with respective R1 leakage ratios reported. All maps are shown in a representative axial slice. The nearest external region (R1) was confined between the RF shield and the outer boundary as indicated by the dashed circle.

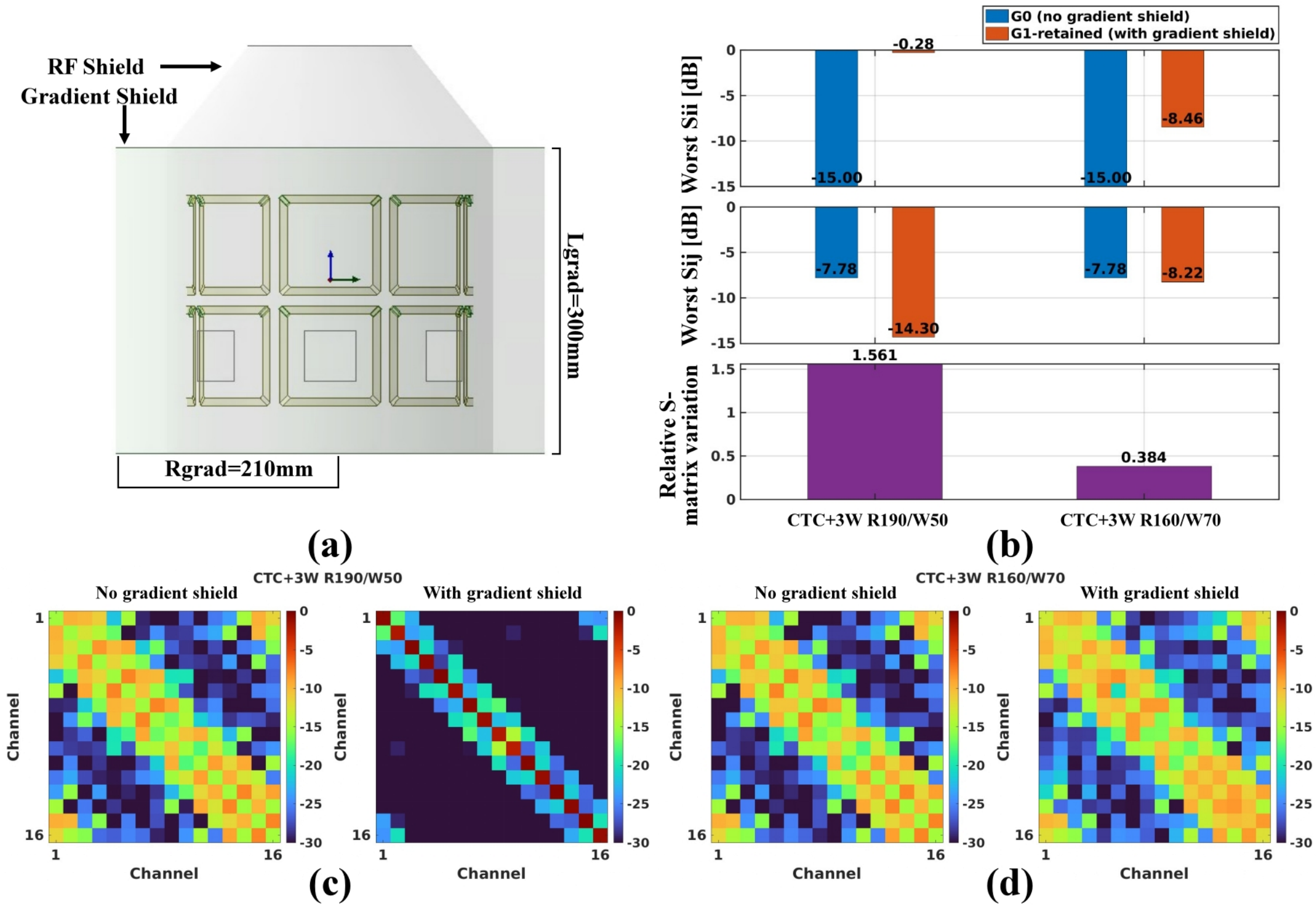

**Fig. 7.** Sensitivity of the two CTC+3W RF-shield configurations to gradient-shield integration. (a) Layout of the gradient shield relative to the shielded transmit array. (b) Worst-channel reflection coefficient (top), worst inter-channel coupling (middle), and relative complex S-matrix variation $D_S$ (bottom) for CTC+3W R190/W50 and R160/W70, comparing the cases without and with the gradient shield. (c), (d) Corresponding S-parameter matrices for CTC+3W R190/W50 and R160/W70, respectively.

### D. Gradient-Shield Integration

The two retained CTC+3W configurations were evaluated for sensitivity to the surrounding gradient shield by including the gradient shield in the EM model while retaining the RF circuit optimized under G0 (Fig. 7).

Under G0, both configurations exhibited worst-channel reflections of approximately -15 dB. After gradient-shield integration (G1-retained), R190/W50 showed a pronounced network perturbation, with the worst-channel reflection degrading to -0.28 dB and the relative complex S-matrix variation increasing to $D_S$ =1.56. By comparison, R160/W70 retained a worst-channel reflection of -8.46 dB and exhibited a substantially smaller S-matrix variation of $D_S$ =0.38. Although the circuit optimization targeted inter-channel coupling below −10 dB, this target was not fully achieved for the retained configurations.

For R160/W70, the worst inter-channel coupling changed only slightly, from -7.78 to -8.22 dB. R190/W50 exhibited lower off-diagonal S-parameter magnitudes after gradient-shield integration; however, these changes occurred concurrently with severe input mismatch and therefore could not be interpreted independently as improved inter-channel isolation.

To characterize the underlying electromagnetic loading, the full 48-port HFSS impedance matrices, including both physical transmit and lumped-element ports, were compared (Fig. 8). For both CTC+3W geometries, the reactive perturbations at the drive and inductive-port locations decreased markedly as the gradient-shield radius increased from 210 to 220–230 mm, indicating a strong dependence on shield-to-gradient spacing. In contrast, the capacitive-port locations exhibited a clear geometry-dependent response. For R160/W70, the reactance changes $|\Delta X_{ii}|$ remained small, whereas R190/W50 retained shifts of approximately 30-45 Ω even at gradient-shield radii of 220-230 mm. Because these differences were obtained directly from the full HFSS impedance matrices, they reflect geometry-dependent electromagnetic loading introduced by the gradient shield rather than differences arising from subsequent circuit optimization.

The corresponding electric-field and surface-current distributions are shown in Supplementary Fig. S5. Compared with R160/W70, R190/W50 exhibited more pronounced field and current redistribution after gradient-shield integration, providing qualitative field-level support for the larger reactive-loading perturbations observed in Fig. 8. The complete port-wise self-reactance changes are provided in Supplementary Fig. S6.

Together, these results indicate that gradient-shield sensitivity depends on both shield-to-gradient spacing and RF-shield geometry, with the latter producing a distinct port-dependent reactive-loading response. Overall, R160/W70 exhibited lower sensitivity to gradient-shield integration and was therefore selected as the final configuration for integrated evaluation.

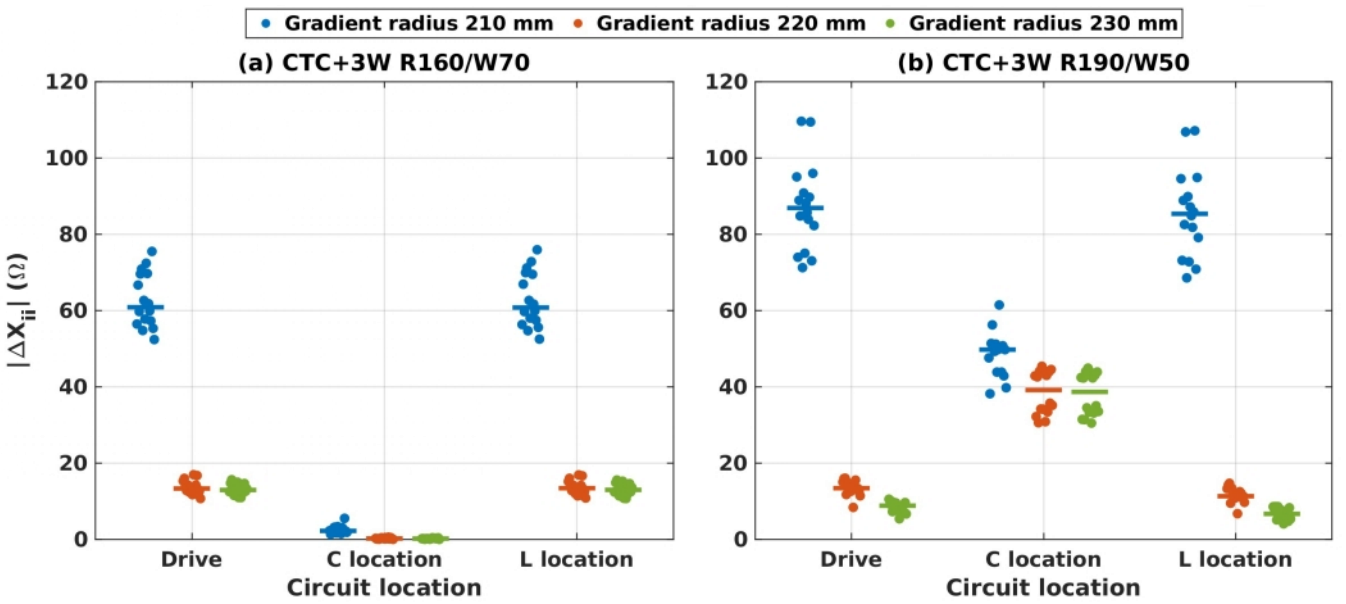

**Fig. 8.** Reactive loading perturbation after gradient-shield integration. Shown are port-wise self-reactance changes, grouped by drive, capacitive (C), and inductive (L) ports, reported for CTC+3W R160/W70 (left) and R190/W50 (right) at gradient-shield radii of 210 (blue), 220 (red), and 230 (green) mm.

### E. Post-Integration RF Performance

The selected array–shield configuration, CTC+3W R160/W70, was further evaluated after inclusion of the gradient shield. Static RF shimming and dynamic pTx were independently optimized using corresponding G0 and G1-retained channel-wise field maps.

Gradient-shield integration resulted in only minor changes in the optimized excitation performance (Fig. 9). The whole-brain static-shim CoV remained nearly unchanged, whereas the dynamic-pTx CoV increased slightly from ~8% under G0 to ~9% under G1-retained. Despite measurable RF-network perturbation, R160/W70 maintained comparable whole-brain excitation performance after gradient-shield integration without RF-circuit retuning.

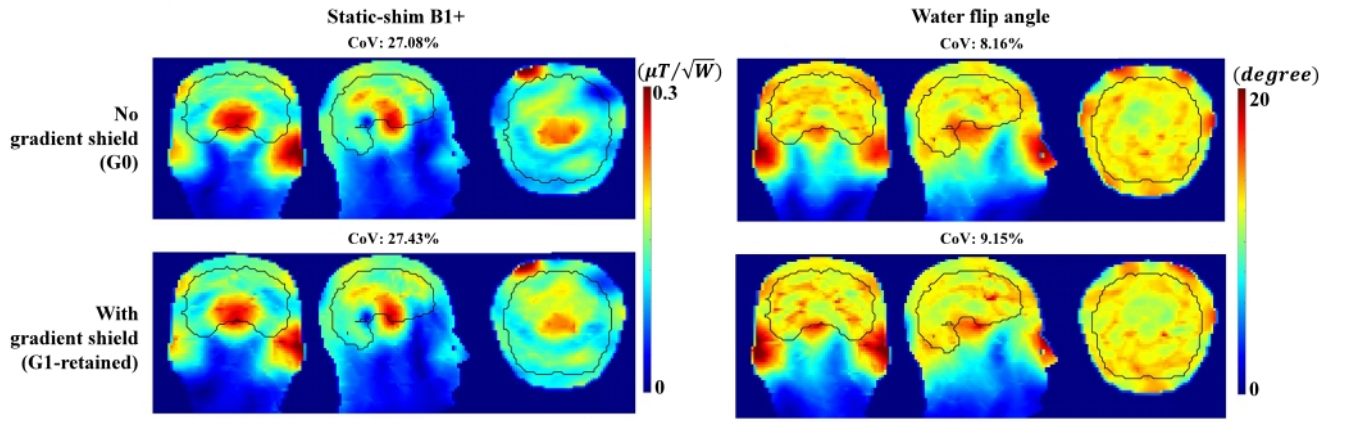

**Fig. 9.** RF performance of CTC+3W R160/W70 before and after gradient-shield integration. Three-view static-shim B1+ distributions (left) and dynamic-pTx water flip-angle distributions (right) are shown without (top) vs with (bottom) gradient shield. Static-shim weights and dynamic-pTx pulses were independently optimized for each condition. Whole-brain CoV is reported.

### F. Experimental Validation

The simulated and measured distributions showed similar spatial patterns across the phantom (Fig. 10), including the central high-efficiency region and the characteristic asymmetry in the coronal plane. Corresponding multislice axial comparisons are provided in Supplementary Fig. S7. The NRMSE between the simulated and measured B1+ efficiency distributions across the phantom was approximately 8%. The observed agreement provides experimental support for the accuracy of the proposed EM/circuit co-simulation framework when used to simulate the transmit RF arrays.

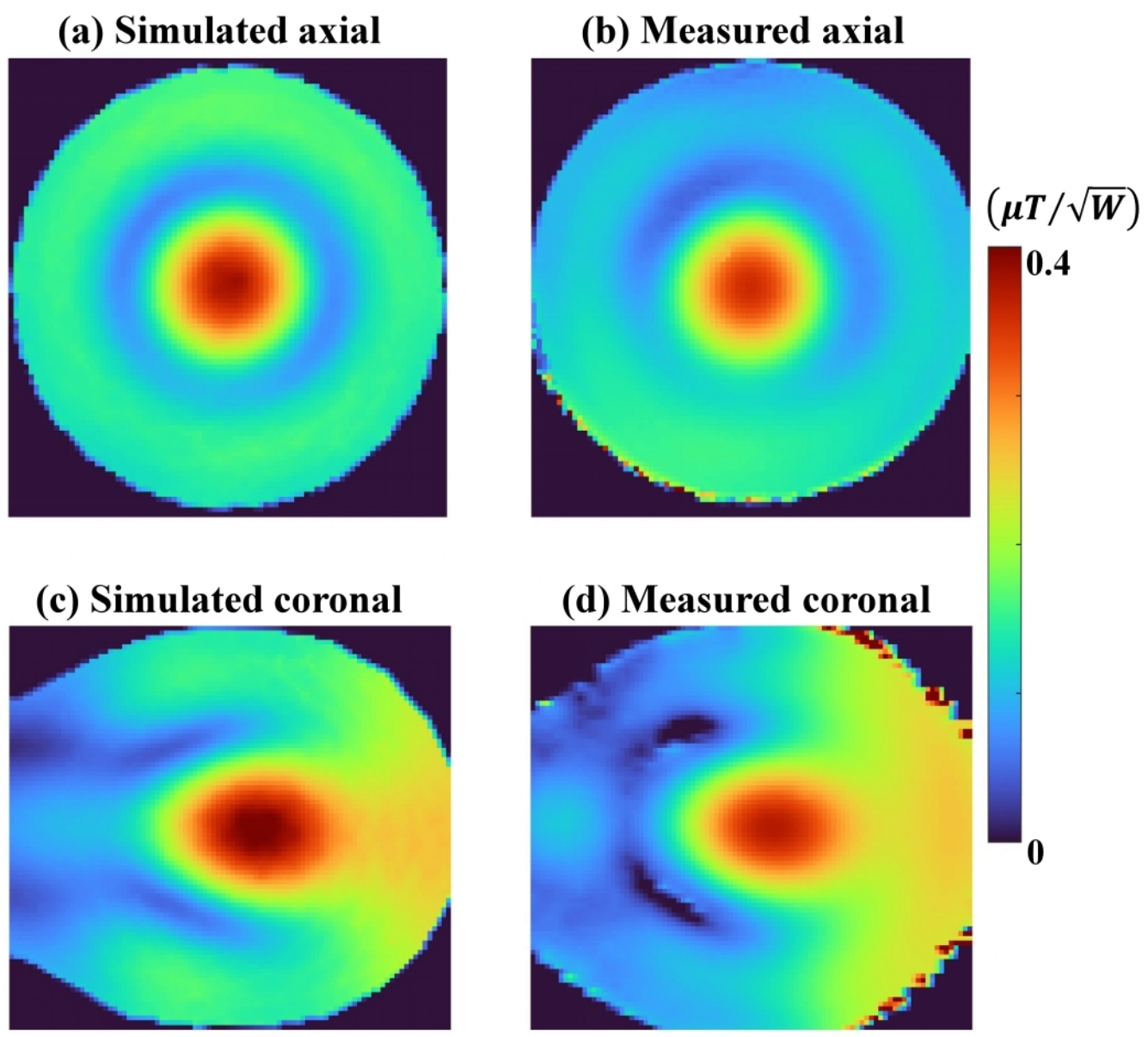

**Fig. 10.** Validating the EM simulation of the unshielded 16-channel LFD transmit RF array using experiments at 10.5 T. Shown are simulated and measured CP-mode B1+ efficiency distributions in representative axial and coronal slices of a uniform PVP lightbulb-shaped phantom. B1+ efficiency was evaluated based on flip angle mapping with $1.5 \times 1.5 \times 5\ \text{mm}^3$ resolution.

## IV. Discussion

The key finding of this study is that local RF-shield geometry constitutes an active electromagnetic design variable that affects both transmit performance and the array's sensitivity to nearby conductive boundaries. The present results show that this integration sensitivity is governed by geometry-dependent reactive loading and cannot be inferred from standalone transmit-performance or leakage metrics alone. Consequently, the shield configuration that performs best under standalone conditions may not necessarily be the most robust configuration after gradient-shield integration.

This behavior was evident in the two retained CTC+3W configurations. R190/W50 provided higher standalone RF efficiency, whereas R160/W70 exhibited lower external-field leakage and substantially smaller gradient-shield-induced network perturbation. The primary integration comparison retained the G0-optimized circuit to directly assess each geometry's sensitivity to the added conductive boundary. A supplementary circuit-retuning analysis showed that much of the mismatch of R190/W50 could be recovered after integration (Supplementary Fig. S8). However, the larger tuning adjustment required for R190/W50 remained consistent with its greater sensitivity to the surrounding conductive boundary.

The full impedance analysis provides an electromagnetic basis for the different integration sensitivities of the two geometries. For both geometries, reactive perturbations at the drive- and inductor-associated ports decreased as the gradient shield was moved outward, showing a clear spacing dependence. At the capacitor-associated ports, however, R190/W50 retained substantially larger reactance shifts than R160/W70 even at increased gradient-shield radii. The corresponding electric-field and surface-current distributions also showed stronger redistribution for R190/W50. Together, these findings indicate that gradient sensitivity depends on shield-to-gradient spacing, local shield geometry, and the resulting port-dependent electromagnetic loading, consistent with the broader sensitivity of high-field transmit arrays to nearby electromagnetic loading structures[39].

Dynamic pTx provided an application-level evaluation of the retained geometries. The ordering in excitation uniformity under static RF shimming was not preserved under dynamic pTx, and the differences among array–shield configurations were reduced after pulse optimization. This behavior is consistent with the additional spatial and temporal degrees of freedom available to dynamic pTx. For the final R160/W70 configuration, optimized whole-brain excitation performance remained comparable after gradient-shield integration while retaining the G0 RF circuit (Fig. 9). A complementary fixed-excitation analysis showed that R190/W50 experienced greater excitation degradation when the G0 shim weights were retained, while circuit retuning partially recovered this loss (Supplementary Fig. S9). The lower network sensitivity of R160/W70 limited the integration-induced degradation, and the remaining field perturbation was compensated during excitation optimization.

The staged screening further showed that RF efficiency, field confinement, and integration robustness represent partially independent design objectives. Previous UHF studies have demonstrated geometry-dependent tradeoffs in transmit-field performance, RF efficiency, and SAR[40], [68], whereas the present results further identify robustness to nearby integrated hardware as an additional design consideration. Localized openings and shield dimensions modified these tradeoffs in a coupled manner across the evaluated configurations. Thus, within the present design space, optimization based on isolated transmit performance alone did not identify the configuration most robust to gradient-shield integration. More broadly, the workflow can be applied to other array geometries or excitation strategies by replacing individual simulation or optimization modules. Alternative excitation approaches, including multi-spoke excitation[52], [53], kT-points[54], and other tailored-pulse methods[55], [56], can be incorporated according to the intended application. The same framework can also be adapted to different excitation targets and array geometries, including partial-brain and open-face designs[44].

The present study also provides a pathway toward implementation in the next-generation 10.5-T head-coil platform. The design findings were used to guide development of the 16-channel transmit array integrated with the second-generation 80-channel receive coil[26]. The experimental CP-mode comparison of the unshielded LFD array showed good agreement between simulated and measured B1+ efficiency distributions, supporting the accuracy of the EM/circuit field predictions for the evaluated array and loading condition. Comprehensive pTx safety validation of the final hardware implementation is addressed separately using the established safety-validation workflow[27].

This study has several limitations. First, the design space was explored through staged screening instead of an exhaustive global search, and the final geometry was selected within the mechanical and system-integration constraints of the intended visual-fMRI and local-gradient application. Second, the interpretation of gradient-induced loading was supported only using the full impedance, electric-field, and surface-current analyses. A more detailed modal analysis could provide further insight into the underlying coupling mechanisms. Third, experimental validation performed here was for the unshielded LFD array. Future work will include validation of the complete RF-shield–gradient-shield configuration once hardware integration is completed. Finally, the evaluated geometries were derived from our first-generation 10.5-T transmit-array architecture[16], [17]. Future studies will investigate how alternative transmit-coil topologies[69], [70], [71] influence RF-shield–gradient interactions.

## V. Conclusion

This work optimized a locally shielded 16-channel transmit array for 10.5-T human brain MRI and its integration with a high-performance head-gradient system. A computational framework was developed for the co-design and system-level evaluation of transmit arrays and RF shields. The results showed that RF-shield geometry strongly influences transmit performance, RF-field confinement, and sensitivity to the surrounding gradient shield through geometry-dependent electromagnetic loading. The final configuration consisted of an LFD array combined with a conical-top cylindrical RF shield with three windows. This design achieved effective RF-field confinement, with R1 leakage below 1%, while exhibiting reduced network perturbation after gradient-shield integration. Without RF-circuit retuning, it also preserved whole-brain excitation performance, with CoVs of approximately 27% under static RF shimming and 9% under dynamic pTx. These findings establish robustness to gradient-shield integration as an important system-level design criterion for locally shielded transmit arrays at ultrahigh field.

## Data and Code availability

The data and MATLAB code will be made publicly available upon publication.